\documentclass[aps,preprint,superscriptaddress,showpacs,nofootinbib]{revtex4}

\usepackage{latexsym}
\usepackage{amssymb}
\usepackage{epsfig,amsmath,graphics}
\usepackage{pstricks}
\usepackage{color}
\usepackage{dcolumn}

\newcommand{\gev}{\text{GeV}}
\newcommand{\mev}{\text{MeV}}

\newcommand{\tev}{\text{TeV}}

\newcommand{\vev}[1]{{\langle #1 \rangle}}

\usepackage{comment}
\newcommand{\be}{\begin{equation}}
\newcommand{\ee}{\end{equation}}

\newcommand{\kev}{{\rm keV}}

\def\bea{\begin{eqnarray}}
\def\eea{\end{eqnarray}}
\def\ltap{\ \raise.3ex\hbox{$<$\kern-.75em\lower1ex\hbox{$\sim$}}\ }
\def\gtap{\ \raise.3ex\hbox{$>$\kern-.75em\lower1ex\hbox{$\sim$}}\ }
\def\lsim{\ \raise.3ex\hbox{$<$\kern-.75em\lower1ex\hbox{$\sim$}}\ }
\def\gsim{\ \raise.3ex\hbox{$>$\kern-.75em\lower1ex\hbox{$\sim$}}\ }

\usepackage{slashed}

\usepackage{color}

\begin{document}

\title{An X-Ray Line from eXciting Dark Matter}
\author{Douglas P. Finkbeiner}
\affiliation{Center for Particle Astrophysics,\\ Harvard University, Cambridge, MA}
\author{Neal Weiner}
\affiliation{Center for Cosmology and Particle Physics, \\Department of Physics, New York University, New York, NY 10003}
\date{\today}
\begin{abstract}
The eXciting Dark Matter (XDM) model was proposed as a mechanism to efficiently convert the kinetic energy (in sufficiently hot environments) of dark matter into e+e- pairs. The standard scenario invokes a doublet of nearly degenerate DM states, and a dark force to mediate a large upscattering cross section between the two. For heavy ($\sim \tev$) DM, the kinetic energy of WIMPs in large (galaxy-sized or larger) halos is capable of producing low-energy positrons. For lighter dark matter, this is kinematically impossible, and the unique observable signature becomes an X-ray line, arising from $\chi \chi \rightarrow \chi^* \chi^*$, followed by $\chi^* \rightarrow \chi \gamma$. This variant of XDM is distinctive from other DM X-ray scenarios in that it tends to be most present in more massive, hotter environments, such as clusters, rather than nearby dwarfs, and has different dependencies from decaying models. We find that it is capable of explaining the recently reported X-ray line at 3.56 keV. For very long lifetimes of the excited state, primordial decays can explain the signal without the presence of upscattering. Thermal models freeze-out as in the normal XDM setup, via annihilations to the light boson $\phi$. For suitable masses the annihilation $\chi \chi \rightarrow \phi \phi$ followed by $\phi \rightarrow SM$ can explain the reported gamma-ray signature from the galactic center. Direct detection is discussed, including the possibility of explaining DAMA via the ``Luminous'' dark matter approach. Quite generally, the proximity of the 3.56 keV line to the energy scale of DAMA motivates a reexamination of electromagnetic explanations. Other signals, including lepton jets and the modification of cores of dwarf galaxies are also considered. 
 \end{abstract}

\maketitle

\section{Introduction}
The evidence for an additional non-baryonic component of the universe has long since been overwhelming. With data from scales as small as dwarf galaxies to the large-scale properties of the universe, from late time evolution of the universe to the properties of the CMB, the Cold Dark Matter (CDM) paradigm has passed numerous tests.

However, although its existence has been well established, its nature remains a total mystery. While a vanilla dark matter particle - devoid of interesting dynamics - remains a viable possibility, dark matter with nontrivial dynamics has been increasingly considered. Attempts to explain direct detection anomalies, indirect detection anomalies, and anomalies in the properties of dwarf galaxy have all relied on invoking a more complicated dark sector, often with multiple dark states and new, dark forces.

One such scenario was the ``eXciting Dark Matter'' (XDM) proposal \cite{Finkbeiner:2007kk}. In it, dark matter $\chi$ is assumed to have an excited state $\chi^*$, and to interact with itself via a new force with mediator $\phi$. Scattering via $\phi$ mediates the excitation $\chi\rightarrow \chi^*$, and the decay $\chi^* \rightarrow \chi + SM$ produces some standard model state, SM.  The idea of XDM was to produce observable signals from the dark matter's {\em kinetic} energy rather than its mass, using the light mediator to produce detectably large scattering cross sections.\footnote{A related scenario with excitations into charged excited states $\chi^\pm$ was considered in \cite{Pospelov:2007xh}.}

For splittings $\delta \equiv m_{\chi^*}-m_\chi \ge 2 m_e$, the decay $\chi^*\rightarrow \chi e^+ e^-$ can be mediated either by kinetic mixing (for a vector mediator) or Higgs portal mixing (for a scalar mediator), and can produce low energy $e^+ e^-$ pairs. Such a scenario might solve one of the oldest mysteries in high-energy astrophysics: the surprising strength of the 511 keV line and associated positronium continuum in the inner galaxy \cite{Knodlseder:2003sv,Churazov:2004as,Knodlseder:2005yq}.

In this work, we explore the $\delta \le 2 m_e$ case, which can only lead to photon or neutrino final states.  The decay $\chi^* \rightarrow \chi+ \gamma$ could produce an X-ray line, and in the following we will consider the detectability of such a signal.

\section{Models}

The original XDM model was based on a simple U(1) dark force with the minimal Lagrangian\footnote{An additional field is assumed to Higgs the dark U(1) at the $100\, \mev \-- 1\, \gev$ scale},

\be
{\cal L} = \bar \chi_i \not \! \! D \chi_i + \frac{1}{4}F^d_{\mu\nu} F^{d\mu\nu}+\epsilon F_{\mu\nu} F^{d\mu\nu}+m^2 \phi_{ \mu} \phi^{\mu}+M_i \bar \chi_i \chi_i+\delta_i \chi_i \chi_i.
\label{eq:lagrangian}
\ee
The kinetic mixing parameter $\epsilon$ gives SM particles $\epsilon$ charge under the dark force, allowing equilibrium in the early universe via $\chi \chi \leftrightarrow \phi \phi$, and $\phi e \leftrightarrow \gamma e$.  

The presence of the excited state allows the possibility of upscattering $\chi \chi \rightarrow \chi^* \chi^*$. For $\delta > 2 m_e$, this will be followed by $\chi^* \rightarrow \chi e^+ e^-$, potentially explaining \cite{Finkbeiner:2007kk,Chen:2009av,Morris:2011dj} the INTEGRAL/SPI positron excess \cite{Knodlseder:2003sv,Churazov:2004as,Knodlseder:2005yq}.

However, for $\delta < 2 m_e$, it has been shown that - absent any other interaction - the excited state is stable on cosmological timescales \cite{Finkbeiner:2009mi,Batell:2009vb}. An obvious modification to the model is the inclusion of a dipole operator $\frac{1}{M}\chi^* \sigma^{\mu \nu} \chi F_{\mu\nu}$, which mediates the decay $\chi^*\rightarrow \chi \gamma$. The lifetime for such a decay is \cite{Chang:2010en}

\be
\tau = 0.5\,  {\rm sec} \times \left(\frac{M}{\tev}\right)^2 \left(\frac{\kev}{\delta}\right)^3.
\label{eq:lifetime}
\ee

Thus, even for $\sim \kev$ splittings, dipoles with $M< 10^8\, \tev$ lead to decays on cosmological timescales. This then motivates us to consider the implications of this XDM scenario for X-ray signals beyond the 511 keV line. We dub this variant of the XDM scenario ``XrayDM''.


\section{X-Ray Signals of XDM and a Feature at 3.56 keV}
While such a model clearly produces X-rays from DM-DM scattering, it is not clear that it produces a {\em detectable} signal of X-rays. To understand whether such a signal is detectable, it is helpful to study this in a specific context.

Recently, \cite{Bulbul:2014sua} reported a potential detection of an X-ray line at 3.56 keV from a stacked combination of clusters, with a particularly bright signal from Perseus. A similar analysis finds a line at the same energy from Perseus and M31 \cite{Boyarsky:2014jta}.
 
The cumulative flux of $\sim 4 \times 10^{-6} {\rm cm^{-2} sec^{-1}}$  from \cite{Bulbul:2014sua} is difficult to interpret as it arises from a combination of clusters at a variety of distances. However, the collaboration does report on the signal from the Perseus cluster individually. Thus, fitting this source gives a simple test as to whether such a signal could arise from XDM.  
Similarly, the analysis of \cite{Boyarsky:2014jta} produces a fit to M31 (and a somewhat broader range of fit to Perseus), giving a second candidate to consider. At the same time, {\em no} signal has arisen from the much closer Virgo cluster, so this limit should be addressed. 

Beginning with Perseus, \cite{Bulbul:2014sua} claims a 3.56 keV line flux of $5.2^{+ 3.70}_{-2.13} \times 10^{-5} {\rm photons \, cm^{-2} \, sec^{-1}}$ (90\% errors) with the cluster core or $2.14^{+1.12}_{-1.05} \times 10^{-5}$ without, arising from XMM MOS observations, and upper limits of $1.77  \times 10^{-5}$ and $1.61  \times 10^{-5}$ for the same regions from the XMM PN observations. The Chandra ACIS-S and ACIS-I observations yielded fluxes of $1.02^{+0.48}_{-0.47}  \times 10^{-5}$ and $1.86^{+1.2}_{-1.6} \times 10^{-5}$.\footnote{While the listed 90\% errors are $+.12$ and $-.16 \times 10^{-5}$, we believe this was a typo, as the errors can be read off from the $\sin^2 \theta$ plot in the paper.} \cite{Boyarsky:2014jta} claim a similar flux of $0.7^{+2.6}_{-2.6} \times 10^{-5}$ (MOS) and $0.92^{+3.1}_{-3.1} \times 10^{-5}$ (PN). 

 For M31 \cite{Boyarsky:2014jta} find a rate of $0.49^{+ 0.16}_{-0.13} \times 10^{-5} {\rm cm^{-2} sec^{-1}}$. For Virgo \cite{Bulbul:2014sua} find an upper limit of $0.91\times 10^{-5} {\rm cm^{-2} sec^{-1}}$.

A naive estimate of the total luminosity from Perseus can be found (assuming an NFW profile) using the cluster parameters found in \cite{SanchezConde:2011ap}
\bea
{\cal L} &=& \int_0^{R_{200}} 4 \pi r^2 \left(\frac{\rho(r)}{m_\chi}\right)^2 \vev{\sigma_{scatt} v} \cr
&=& 1.9 \times 10^{49} {\rm photons/sec} \times \left(\frac{\vev{\sigma_{scatt} v}}{10^{-19} {\rm cm^3 sec^{-1}}}\right)\left(\frac{10 \gev}{m_\chi}\right)^2.
\eea
With Perseus 78 Mpc away, this yields a local photon flux
\bea
\Phi = 2.6 \times 10^{-5} \left(\frac{\vev{\sigma v}}{10^{-19} {\rm cm^3 sec^{-1}}}\right) \left(\frac{10 \gev}{m_\chi}\right)^2{\rm photons/sec}.
\eea

Clearly, this cross section is well above the conventional thermal annihilation cross section, but since this is a {\em scattering} process, this cross section can be naturally large, as we now describe.

The perturbative cross section for this scattering has a cross section
\be
\sigma = \frac{4 \pi M_\chi^2 \alpha_{d}^2}{m_\phi^4},
\ee
where $\alpha_d$ is the dark U(1) coupling constant. For $\alpha M_\chi /m_\phi \sim 1$ one must worry about non-perturbative effects and appropriately resum the ladder diagrams, in which case a numerical calculation becomes motivated \cite{Morris:2011dj}.

Nonetheless, such a cross section is easily achievable. The unitarity bound on the S-wave component of the cross section is $\sigma v \sim \pi/(M_\chi^2 v)\sim 10^{-16} \rm cm^{3}\, sec^{-1}$, for cluster velocities of $\sim .003 c$. 
Moreover, for light mediators and glancing collisions, the scattering is naturally composed of multiple partial waves \cite{Morris:2011dj}, yielding often an approximately geometric cross section, i.e., $1/q^2$ or $1/m_\phi^2$ for $m_\phi^2 > q^2$.

Thus, for $m_\phi \sim 100 \mev$, we expect a cross section saturating at levels as large as $4 \times 10^{-26} \rm cm^2$ yielding  $\sigma v \sim 10^{-17}\rm cm^3 sec^{-1}$ for relative velocities of $2 \times 1000 \, {\rm km/sec}$.  Clearly, achieving this level of {\em rate} is possible. 
\subsection{Estimating Rates}
We now go about more precisely attempting to address the signals observed. While a detailed fit to the full stacked cluster analysis is beyond our scope, we can reasonably attempt to understand the Chandra observation of the line in Perseus, M31 and the non-observation in Virgo.

While the NFW model is a simple and convenient parametrization, it is thought that different histories can lead to different halo profiles \cite{Graham:2006ae}. In particular, Virgo is thought to be a younger cluster, and may not have fully settled into a steep profile as yet. To study the effects on the signal, we use a variant of the NFW profile
\be
\rho(r) = \frac{\rho_0}{(r/R_s)^\gamma (1+r/R_s)^{3-\gamma}},
\ee 
where we hold the total mass inside the virial radius fixed by varying $\rho_0$. 

To compare signals, we approximate the Chandra ACIS-I FOV (which is a $2 \times 2$ array of $8' \times 8'$ CCDs) by a $9'$ disc, which has nearly the same angular size. For cuspy profiles (as will be necessary to explain the data), the majority of the signal is in the central region, thus the precise boundary is not important at the leading order. For M31, whose flux has been found by \cite{Boyarsky:2014jta} from XMM data, we take a radius around the inner 15' as our region.

We will parametrize the predicted flux as
\be
\Phi_{perseus} = F^{19,10}_{perseus} \times \left(\frac{\langle\sigma v \rangle_{perseus}}{10^{-19}{\rm cm^3 \, sec^{-1}}}\right) \left(\frac{10 \gev}{m_\chi}\right)^2.
\ee

For inner slope profiles $\gamma = (0.7,1,1.3)$ we find $F^{19,10}_{perseus}= (1.1,2.0,5.9) \times 10^{-5} \rm cm^{-2}\, sec^{-1}$. While for Virgo, we find $F^{19,10}_{virgo}= (2.7, 9.6,62) \times 10^{-5} \rm cm^{-2}\, sec^{-1}$. Finally, for M31, $F^{19,10}_{M31}= 5\times (1.2,3.7,18) \times 10^{-5} \rm cm^{-2}\, sec^{-1}$. Note that for M31, the characteristic velocity is $\sim \times 5$ lower than in Perseus, and we pull out an overall factor of 5 to make the comparison clearer.

\subsection{More detailed calculation}
The above calculation assumed that $\vev{\sigma_{scatt} v}$ is independent of location in the cluster, and has a fixed value in each system.  As a more careful estimate, we now take
\be
\vev{\sigma_{scatt} v} = \sigma_{mr} \sqrt{v^2-v_{thresh}^2},
\ee
where $\sigma_{mr}$ is the cross section in the ``moderately relativistic'' limit, and $v$ is the relative velocity of the WIMPs.  We take the (3D, single-particle) rms velocity dispersion to be $\sqrt{3/2} v_{circ}$, where $v_{circ}$ is the circular velocity at the scale radius.  At each radius we truncate the velocity distribution at the escape velocity, $v_{esc}(r)$.  

Taking a reference value of $\sigma_{mr} = 10^{-28} \rm{cm}^2$, we get
$F_{perseus} = (0.12,0.29,1.1)\times10^{-5}$ in a 9 arcmin radius for (0.7,1.0,1.3). 
$F_{virgo} = (0.47,2.0,13.0)\times10^{-5}$ in a 9 arcmin radius for (0.7,1.0,1.3). 
$F_{M31} = (0.29, 1.3, 9.6)\times10^{-5}$ in a 15 arcmin radius. 
As we see, the variation with the slope $\gamma$ is even more pronounced for this model.  But the picture is qualitatively the same as the naive model.

As a result, we see that for ``pure'' NFW profiles, there seems to be a conflict between the non-observation in Virgo and the detection in Perseus. However, because the upscattering process is proportional to $\rho^2$, there is a significant dependence on the halo, and moderate variations away from NFW can easily make these results consistent. More colloquiually, the fact that we see Perseus and not Virgo may just be a measurement of their dark matter profiles.

For M31, assuming that Perseus is relatively steep, we would expect a signal at a similar order, but potentially larger or smaller. The claim of \cite{Boyarsky:2014jta} for a similar flux is consistent with this.
Consequently, we find that DM-DM scattering can explain the presence of an X-ray line in Perseus, consistent with non-observation in Virgo. Moreover, a comparable signal is possible in M31.

Up to this point, we have assumed that the X-ray signal traces the scattering. For short lifetimes of the excited state this is true, but for longer lifetimes this would not be. For large dipoles, the excited WIMP decays rapidly and the X-ray signal traces the $\rho^2$ profile. In contrast, when the lifetime is long compared to the dynamical time of the cluster, the distribution will trace the (scattered) WIMP distribution. To the extent that glancing collisions dominate this process, this will be similar to the standard WIMP distribution. Consequently, when varying the lifetime, the potential signal interpolates between $\rho$ and $\rho^2$.

\subsection{Signals from a Primordial Excitation}
With a lifetime shorter than the age of the universe, this scenario relies on active excitations in order to produce an observable 3.56 keV line. However, it is also possible that we are simply viewing the decays of a primordial excited population.

As noted in \cite{Finkbeiner:2008gw}, the decays of a relic population of excited states can have interesting astrophysical signals. With $\delta \sim \mev$, fractions as large as $\sim 0.01$ are possible. As discussed in \cite{Finkbeiner:2009mi,Batell:2009vb}, it is quite natural to have a larger excitation fraction in XDM models with smaller splittings. Downscattering processes off of electrons naturally decouple early (compared to keV) for $\delta\sim 3.5 \kev $ and assuming $m_\phi \sim \gev$ and $\epsilon \sim 10^{-4}$. Self-scattering decouples as $\chi^*\chi^* \rightarrow \chi \chi$ gets suppressed below $T< 3.56 \kev$ and the number density of $\chi+\chi^*$ is already small. Nonetheless, assuming that kinetic decoupling occurs at $T\approx m_e$, one expects a primordial fraction $f \gsim 10^{-3}$.

Normalizing to a sterile neutrino signal for convenience, we can use the lifetime in eq. \ref{eq:lifetime} to determine the observed signal. For an excited fraction $f$, the dipole scale $M$ that will give an equivalent signal to a sterile neutrino is
\be
M_{dipole} = 5 \cdot 10^{10}\, \tev \times \sqrt{\frac{f}{10^{-3}}}\sqrt{\frac{10^{-10}}{\sin^2 2 \theta}}\sqrt{\frac{\gev}{ m_\chi}} \left(\frac{7.12 \kev}{m_s}\right)^2 \left(\frac{\delta}{3.56 \kev}\right)^{3/2}.
\ee

Thus, even for situations where the scattering is insufficient to provide the rate, decays of the primordial excited state could give the comparable signal to the sterile neutrino case.

\section{Other Signals}

\subsection{Direct Detection}
The direct detection signals of such a scenario vary depending on the mass and lifetime of the excited state. DM-nucleus scattering can occur either via kinetic mixing with the photon, or via the dipole. 

For a very massive state ($m_\chi \sim 100 \gev$) the inelasticity is essentially irrelevant, and $\chi$ will appear as standard, elastically scattering DM, mediated by the vector or dipole interaction.

For a lighter state ($m_\chi \lsim 10 \gev$), and suitably large vector interaction, this becomes a realization of inelastic dark matter \cite{TuckerSmith:2001hy}, with the usual features such as peaked spectra, amplified modulation, and the preference to scatter off of heavy targets. Even with a fairly large dipole (e.g., TeV-suppressed or weaker), the lifetime of the excited state will be easily long enough for it to escape detectors. For larger dipoles, this would instead appear as a variant of magnetic inelastic dark matter \cite{Chang:2010en}.

However, if both the vector interaction and dipole yield scatterings too rare to be detectable, but the dipole is large enough to mediate decays, we arrive at a possible variant of the ``luminous'' dark matter scenario \cite{Feldstein:2010su}. In such a scenario, dark matter can upscatter in rock, then decay electromagnetically in a detector, and was proposed as an explanation of the DAMA modulation. Indeed, a search at CUORE for precisely this hybrid (vector+dipole) scenario has recently been proposed \cite{Pospelov:2013nea}. Taking the recent X-ray cluster results for normalization, this gives a specific target to look at, both as an anomalous peak at 3.56 keV in the bolometer spectrum as well as for modulation in that peak.

For such a model to explain DAMA seems challenging, however, as the peak produced at $\sim 3.56 \kev$ is far from the observed peak, which is closer to the $^{40}$K peak at 3.2 keV. That said, the proximity of the energies of these two anomalies motivates a closer examination of all assumptions.  The 3.2 keV Auger electron/X-ray signal used for calibration is seen in coincidence with a 1.46 MeV gamma in a neighboring crystal.  If the gamma ray were to deposit even 0.02\% of its energy before leaving the crystal containing the decay, this could introduce a 10\% shift in the calibration of the DAMA energy scale.  

\subsection{Galactic Center}
While the original motivation for the light mediator is to allow large WIMP upscattering, it can also show up in higher energy cosmic ray signals, arising from $\chi \chi \rightarrow \phi \phi$ followed by $\phi \rightarrow SM$ \cite{Finkbeiner:2007kk,Pospelov:2007mp,Cholis:2008vb, TODM,Pospelov:2008jd,Nomura:2008ru}. In this particular case, for a $\sim 10 \gev$ WIMP annihilating into a $\phi$ with mass in the $\sim 100\mev \-- \gev$ range, we can expect signals in the sub-10 GeV range.

Interestingly, such a signal has been claimed in the galactic center \cite{Goodenough:2009gk,Hooper:2010mq,Hooper:2010im,Hooper:2011ti,Abazajian:2012pn,Hooper:2013rwa,Abazajian:2014fta,Daylan:2014rsa}. As shown in \cite{Hooper:2012cw}, dark matter annihilating via a light mediator can fit these data well for an approximately thermal cross section. Thus, the combination of these signals (X-ray and gamma-ray) motivate additional low energy searches for the mediator $\phi$ \cite{Essig:2009nc,Bjorken:2009mm,Essig:2010xa,Wojtsekhowski:2009vz}, such as those at APEX \cite{Essig:2010xa,Abrahamyan:2011gv}, MAMI \cite{Merkel:2011ze}, HPS \cite{HPS} and DarkLight \cite{Freytsis:2009bh}, searches at low energy$e^+e^-$ colliders \cite{AmelinoCamelia:2010me,Archilli:2011nh,Aubert:2009af,Lees:2012ra,Jaegle:2012sv}  as well LHC searches \cite{Chatrchyan:2011hr,Aad:2012qua} for the associated lepton jets signal \cite{ArkaniHamed:2008qp,Baumgart:2009tn}. (For a review, see \cite{Essig:2013lka}.)

\subsection{Cored Profiles for Dwarf Galaxies}
As is well known, a large self-interaction cross section can alter the properties of galaxies and galaxy clusters \cite{2000PhRvL..84.3760S}. Such effects could be desirable in explaining potential deviations from CDM that have been recently observed \cite{BoylanKolchin:2011de,BoylanKolchin:2011dk,Strigari:2011ps}. Nonetheless, there are strong constraints on such large self-interactions \cite{Feng:2009hw,Buckley:2009in}. The velocity dependences of the cross section produced from a Yukawa potential can ameliorate a number of constraints, while still having the relevant effects for dwarf galaxies \cite{Loeb:2010gj,Rocha:2012jg,Tulin:2012wi,Tulin:2013teo}. While the presence of the excited state will certainly modify the scattering, to the extent that it is only slightly off-shell, large self-interaction cross sections are still expected. 

Complicating matters is that in small halos where excitations {\em can} occur, the production of X-rays can add pressure on the baryonic gas, altering star formation.

While a quantitative study is beyond our scope, it is worth recognizing that the interesting properties for dwarf galaxies from dark force models would be expected to persist in the presence of a small splitting to the excited state. Moreover, the production of X-rays may alter star formation in these halos in meaningful ways.

\section{Discussion}
The idea that the mass of dark matter can be converted into a cosmic ray signal has long inspired numerous searches, leveraging its high energy to compete with backgrounds. In XDM, the cosmic ray signal instead arises by converting the {\em kinetic} energy of dark matter into a cosmic ray signal, but now leveraging a light dark force carrier to yield a large {\em scattering} cross section, converting dark matter $\chi$ into its excited state $\chi^*$. For  $\delta = m_{\chi^*}-m_{\chi}< 2 m_e$, the natural signal is that of an X-ray line, from $\chi^* \rightarrow \chi + \gamma$. We have shown here that such a signal is observable, and, moreover, is compatible with recent claims of a 3.56 keV line from clusters and M31. In cases with very suppressed dipoles, a primordial excited fraction of dark matter can explain the signal.

Obviously, as a DM explanation of the 3.56 keV line, the scenario is more involved than that of a simple sterile neutrino. Nonetheless, there are good reasons to pursue a second line of thinking. To begin with, the spatial morphology would allow a comparison (with the XDM signal falling more rapidly with radius), and the $\rho^2$ dependence might alleviate some tensions present with the sterile neutrino model, in particular with Virgo.

The XDM scenario also offers different search strategies - for instance, one could stack a large set X-ray observations of nearby ellipticals selected from the SDSS catalog. Finally, one can look for XDM signals in direct detection experiments, as well as at accelerators in searches for the force carrier $\phi$. These final strategies are relevant when primordial excited state decay explains the line. Should this signal persist, we then have great hope that additional observations could clarify what is the origin of the 3.56 keV line.

\section*{Acknowledgements}
We thank Jia Liu and Natalia Toro for useful discussions. DF is partially supported by the NASA Fermi Guest Investigator Program. NW is supported by NSF grants PHY-0947827 and PHY-1316753.

\bibliography{XrayDM}
\bibliographystyle{apsrev}

\end{document}